\documentclass[aps,prl,twocolumn,superscriptaddress,groupedaddress]{revtex4}  % for review and submission
\usepackage{graphicx}  % needed for figures
\usepackage{dcolumn}   % needed for some tables
\usepackage{bm}        % for math
\usepackage{amssymb}   % for math
\usepackage{soul,ulem}   % for math

\begin{document}

% The following information is for internal review, please remove them for submission
%\widetext
%\leftline{Version xx as of \today}
%\leftline{Primary authors: Joe E. Physics}
%\leftline{To be submitted to (PRL, PRD-RC, PRD, PLB; choose one.)}
%\leftline{Comment to {\tt d0-run2eb-nnn@fnal.gov} by xxx, yyy}
%\centerline{\em D\O\ INTERNAL DOCUMENT -- NOT FOR PUBLIC DISTRIBUTION}

% the following line is for submission, including submission to the arXiv!!
%\hspace{5.2in} \mbox{Fermilab-Pub-04/xxx-E}

\title{Tailoring correlations of the local density of states in disordered photonic materials}
%\input author_list.tex       % D0 authors (remove the first 3 lines
                             % of this file prior to submission, they
                             % contain a time stamp for the authorlist)
                             % (includes institutions and visitors)
\author{F. Riboli}
\email{francesco.riboli@ino.it}
%%\email{francesco.riboli@gmail.com}
\affiliation{Department of Physics, University of Trento, via Sommarive 14, 38050 Povo (TN), Italy}%
\affiliation{Istituto Nazionale di Ottica, CNR, S.S. Sesto Fiorentino, 50019 Sesto Fiorentino, Italy}%
\affiliation{European Laboratory for Nonlinear Spectroscopy, via Nello Carrara 1, Sesto Fiorentino, FI, Italy}%
\author{F. Uccheddu}%
\affiliation{Department of Industrial Engineering, University of Florence, via Santa Marta 5, 50139 Firenze, Italy}%
\author{G. Monaco}%
\affiliation{Department of Physics, University of Trento, via Sommarive 14, 38050 Povo (TN), Italy}%
\author{N. Caselli}%
\affiliation{European Laboratory for Nonlinear Spectroscopy, via Nello Carrara 1, Sesto Fiorentino, FI, Italy}%
\affiliation{Department of Physics, University of Florence, Via G. Sansone 1, 50019 Sesto Fiorentino (FI), Italy}%
\affiliation{Instituto de Ciencia de Materiales de Madrid, CSIC, Calle Sor Juana Inés de la Cruz, 3, Madrid, Spain}%
\author{F. Intonti}%
\affiliation{European Laboratory for Nonlinear Spectroscopy, via Nello Carrara 1, Sesto Fiorentino, FI, Italy}%
\affiliation{Department of Physics, University of Florence, Via G. Sansone 1, 50019 Sesto Fiorentino (FI), Italy}%
\author{M. Gurioli}%
\affiliation{European Laboratory for Nonlinear Spectroscopy, via Nello Carrara 1, Sesto Fiorentino, FI, Italy}%
\affiliation{Department of Physics, University of Florence, Via G. Sansone 1, 50019 Sesto Fiorentino (FI), Italy}%
\author{S.E. Skipetrov}%
\email{Sergey.Skipetrov@lpmmc.cnrs.fr}
\affiliation{Universit\'{e} Grenoble Alpes, LPMMC, F-38000 Grenoble, France}%
\affiliation{CNRS, LPMMC, F-38000 Grenoble, France}%

\date{\today}

\begin{abstract}
We present experimental evidence for the different mechanisms driving the fluctuations of the local density of states (LDOS) in disordered photonic systems. We establish a clear link between the microscopic structure of the material and the frequency correlation function of LDOS accessed by a near-field hyperspectral imaging technique.  We show, in particular, that short- and long-range frequency correlations of LDOS are controlled by different physical processes (multiple or single scattering processes, respectively) that can be---to some extent---manipulated independently. We also demonstrate that the single scattering contribution to LDOS fluctuations is sensitive to subwavelength features of the material and, in particular, to the correlation length of its dielectric function. Our work paves a way towards a complete control of statistical properties of disordered photonic systems, allowing for designing materials with predefined correlations of LDOS.
\end{abstract}

%\pacs{}
\maketitle

%\section{\label{sec:level1}First-level heading}
% sections are not used for PRL papers

After more than a hundred years of intense research on light propagation in random media, we now start to realize that disorder is not only a nuisance for imaging and telecommunications but that it can be exploited to design new functional materials outperforming ``clean''  systems in a number of applications \cite{Polman, Vynck, Sapienza, Wiersma, Cao1, Cao2, Pappu, Goorden}. However, designing an efficient disordered photonic material requires controlling the statistics of its optical properties. Such a control has been already achieved, to a large extent, for transport properties governing propagation of light (scattering and transport mean free paths, diffusion coefficient, etc. \cite{Akkermans}) but remains only partial for the properties relevant for the emission of light. The latter is a complicated process \cite{Faez} but in many situations its efficiency, as well as absorption efficiency and many other types of light-matter interaction, depend on the local density of states (LDOS) at the source position  \cite{Loudon}. LDOS $\rho ( \mathbf{r},\nu )$ is simply a number of optical states (modes) at a point $\mathbf{r}$ and at a frequency $\nu$, per unit volume and unit frequency band. In a disordered material, LDOS fluctuates in space and with the frequency of light \cite{Akkermans} as demonstrated in recent experiments \cite{Birowosuto,Carminati,Sapienza2,Garcia}. Fluctuations of LDOS at the source position lead to fluctuations in the decay rate of spontaneous emission \cite{Loudon} and produce long-range spatial correlations of emitted intensity in the far field \cite{Skipetrov}.

\begin{figure}
\includegraphics[width=\columnwidth]{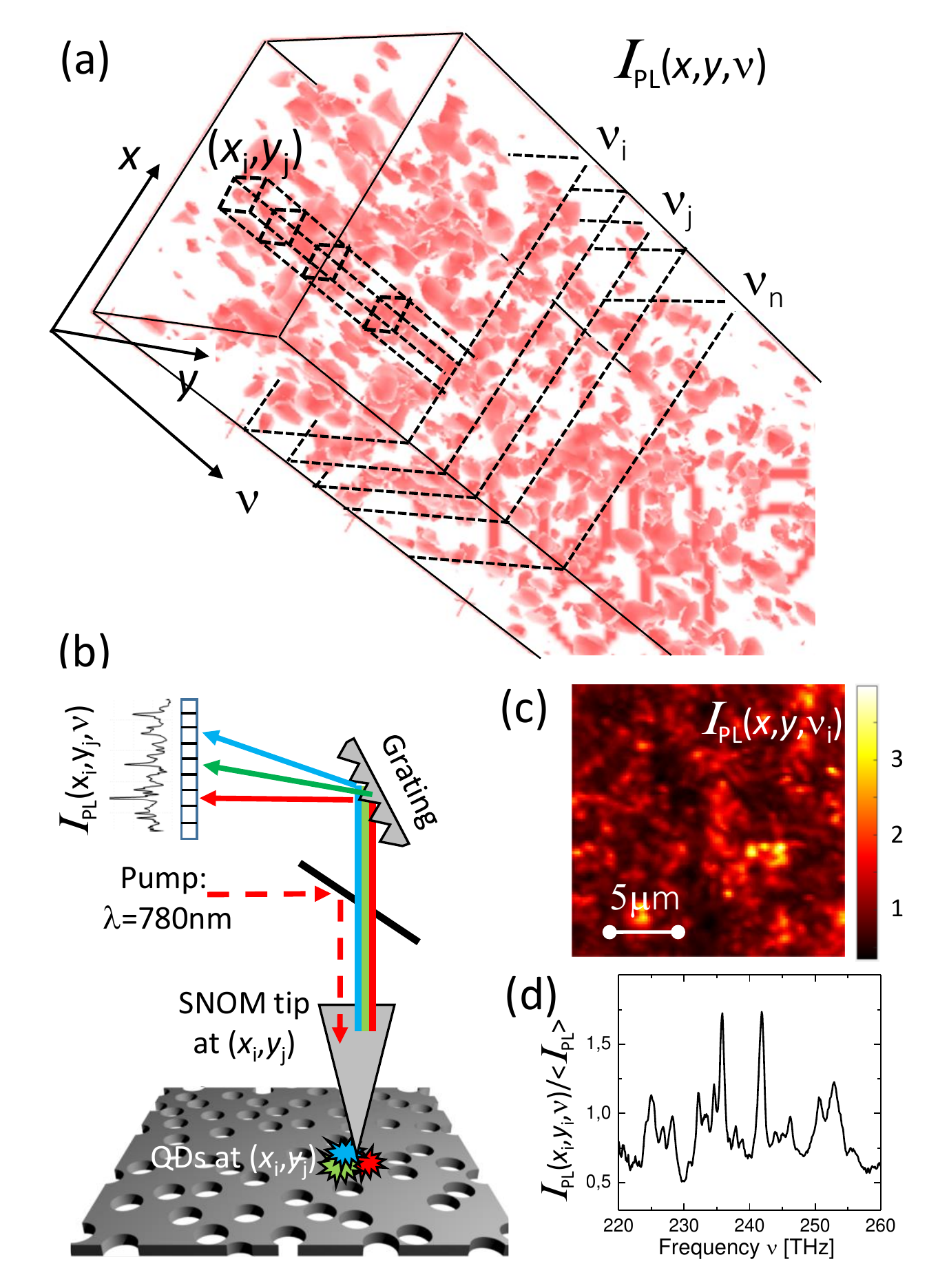}
\caption{\label{fig1} Near-field hyperspectral imaging of QD photoluminescence. (a) 3D equi-intensity surface plot of PL signal $I_{\mathrm{PL}}( \mathbf{r},\nu )$ in a typical experimental scan. The total number of voxels of the 3D image is of the order of $4\times 10^{6}$. (b) Sketch of the experiment. (c) and (d) show PL intensity as a function of position for a given, randomly chosen frequency $\nu_{i}$,  and as a function of frequency for a given, randomly chosen position $(x_{i},y_{i})$, respectively.}
\end{figure}

\begin{figure}
\includegraphics[width=\columnwidth]{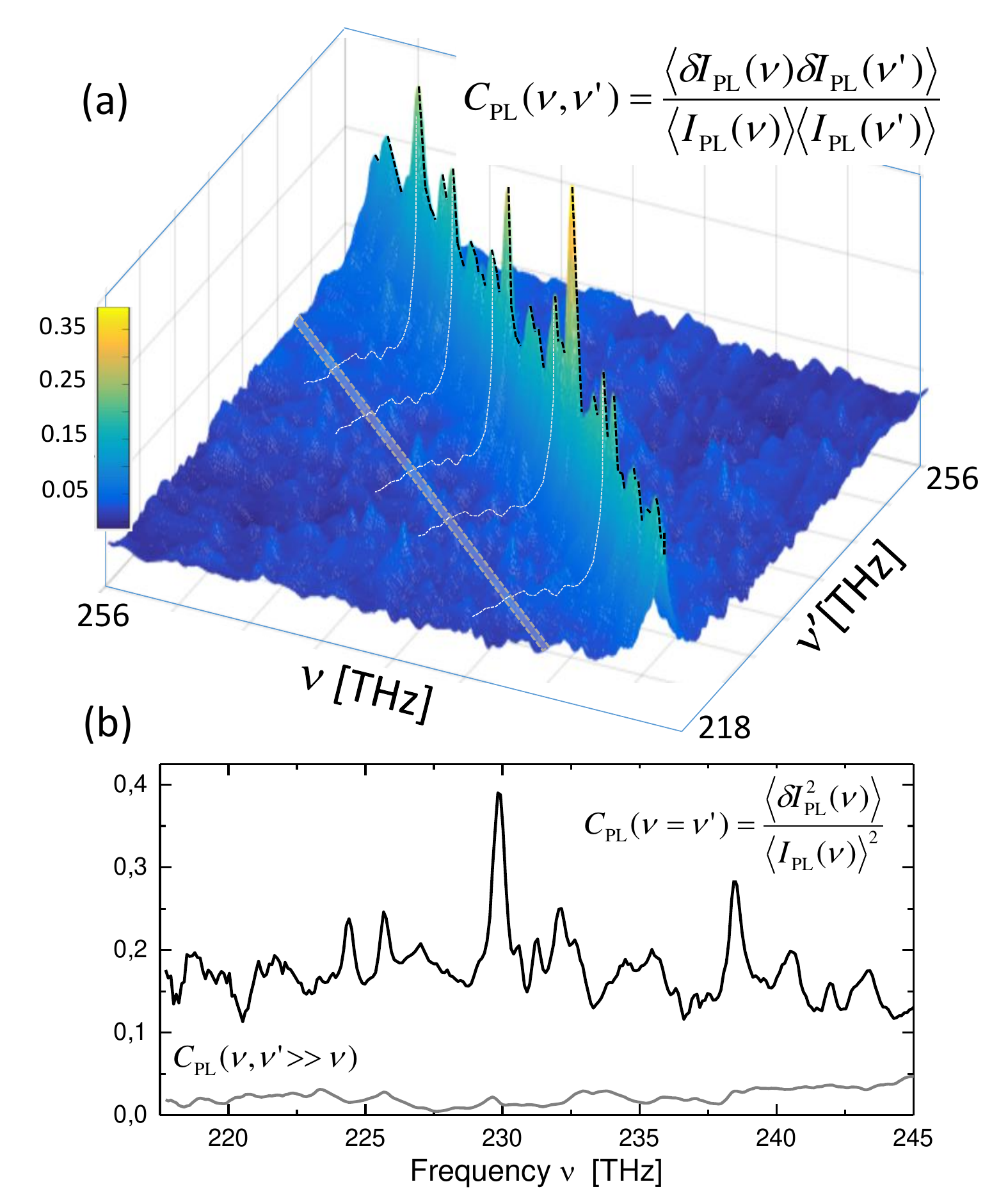}
\caption{\label{fig2} Frequency-resolved correlation analysis. (a) Frequency correlation matrix of QD photoluminescence for a system with $k\ell^{*} = 4$, which in terms of structural parameters corresponds to an average hole diameter $\langle d_{h} \rangle = 210$ nm and a hole surface filling fraction $f = 0.35$. (b) The diagonal elements of $C_{\mathrm{PL}}(\nu,\nu')$ equal to the normalized variance of PL intensity fluctuations (upper curve). The gray lower curve shows the fluctuations of off-diagonal terms ($\nu \ne \nu'$) at large detuning $\nu'\gg\nu$. It was evaluated along the gray line in the panel (a).}
\end{figure}

Here we probe LDOS statistics using the near-field hyperspectral imaging technique \cite{Riboli}. Our experiments probe photoluminescence (PL) of InAs quantum-dots (QDs) embedded in dielectric (GaAs) planar waveguides. Disorder is realized by perforating the waveguides with randomly distributed circular holes \cite{Riboli,SM}. The QDs are excited through a dielectric tip of a near-field optical microscope (SNOM) with a low-power diode laser. PL of QDs is collected through the same tip [see Fig.~\ref{fig1}(b)]. The measured PL intensity $I_{\mathrm{PL}}( \mathbf{r},\nu )$ is recorded every $200$ nm on a square spatial grid. As we show in Fig.~\ref{fig1}(a), $I_{\mathrm{PL}}( \mathbf{r},\nu )$ exhibits strong fluctuations with both the position of the SNOM tip  $\mathbf{r} = (x,y)$ [Fig.~\ref{fig1}(c)] and frequency $\nu$ [Fig.~\ref{fig1}(d)]. A typical set of data for one sample comprises a region of interest of 18 $\mu$m $\times$ 18 $\mu$m, centered in the middle of the sample, far from the boundaries. For each position of the SNOM tip we collect PL signal between 218 THz and 260 THz with a frequency resolution of 0.1 THz. The fluctuations of PL intensity are characterized by an intensity correlation matrix
\begin{equation}
 C_{\mathrm{PL}}(\nu,\nu')=\frac{\langle   \delta I_{\mathrm{PL}}(\mathbf{r},\nu) \delta I_{\mathrm{PL}}(\mathbf{r},\nu')\rangle}{\langle  I_{\mathrm{PL}}(\mathbf{r},\nu)\rangle \langle I_{\mathrm{PL}}(\mathbf{r},\nu')\rangle},
\label{eqn1}
\end{equation}
where $\delta I_{\mathrm{PL}}(\mathbf{r},\nu) = I_{\mathrm{PL}}(\mathbf{r},\nu) - \langle I_{\mathrm{PL}}(\mathbf{r},\nu) \rangle $. The averaging $\langle \textellipsis \rangle$ is performed over the region of interest. Each element of the matrix $C_{\mathrm{PL}}(\nu,\nu')$ is an average of $8 \times 10^{3}$ correlated values. Normalization by the average PL intensities in Eq.\ (\ref{eqn1}) minimizes the influence of the intrinsic structure of QD emission spectrum (i.e. a spectrum that would be measured in the absence of disorder). Figure \ref{fig2}(a) shows the typical correlation matrix $C_{\mathrm{PL}}(\nu,\nu')$ for a sample with $k\ell^{*} = 4$, where $k = (2\pi/\lambda)n_{\mathrm{eff}}$ effective wavenumber of light in the sample, $n_{\mathrm{eff}}$ is the effective refractive index, and $\ell^{*}$ is the transport mean free path \cite{Akkermans}. We observe strong variations of $C_{\mathrm{PL}}(\nu,\nu')$ with frequencies. The variations are particularly pronounced for the diagonal elements $\nu = \nu'$ and are weaker for off-diagonal elements but persist even at large detunings $|\Delta\nu|=|\nu-\nu'|$. These variations are a combination of the intrinsic fluctuations of the system's parameters in space, residual statistical fluctuations due to a finite size of the statistical ensemble, and other extrinsic effects. The contribution of the latter is estimated to be below $20\%$ of the overall signal variation \cite{SM}. The autocorrelation function $C_{\mathrm{PL}}(\Delta \nu)$ of the signal is obtained by averaging the correlation matrix over $\nu$ and $\nu'$ at a constant detuning $\Delta\nu$. This frequency averaging further decreases the contribution of extrinsic effects and allows for comparing experimental data $C_{\mathrm{PL}}(\Delta \nu)$ with theory.

In general, the relation between PL intensity due to QDs embedded in a disordered sample and radiative LDOS is not trivial. However, as discussed in Ref.\ \cite{Intonti_PRB} and further in the Supplemental Material \cite{SM}, for our samples, a linear relation can be established between PL intensity and the local density of states having the electric field component in the sample plane. For brevity, we abbreviate the latter quantity as LDOS in the following, although one has to understand that it represents only one of the contributions to the total LDOS. A linear relation between PL and LDOS accounts for roughly 80\% of the measured signal \cite{SM}.
Our calculation of the correlation function of LDOS integrated over a measurement area $S$,  $\langle C_{\rho}(\Delta \nu)\rangle$ ---a quantity that can be directly compared to $C_{\mathrm{PL}}(\Delta \nu)$---is described in the Supplemental Material \cite{SM}. The result is a sum of an infinite-range (not decaying with $\Delta\nu$ as far as $|\Delta\nu|\ll\nu$) and short-range (rapidly decaying with $\Delta\nu$) contributions:
\begin{eqnarray}
\langle C_{\rho}(\Delta\nu)\rangle &=& F_{1}(k\ell_{\epsilon},ka,k\ell)\frac{\ln(2k\ell)}{\pi k\ell}
\nonumber \\
&+& F_{2}(ka)\mathrm{Re}\left[\frac{D_{B}}{D(\Delta\nu)}-1\right],
\label{eqn2}
\end{eqnarray}
where $\ell$ is the in-plane scattering mean free path \cite{Akkermans}, $a$ is the radius of the signal collection area $S$ assumed circular. The renormalized in-plane diffusion coefficient $D(\Delta\nu)$ obeys \cite{SM,Gorkov}
\begin{equation}
\frac{D(\Delta\nu)}{D_{B}}=1-\frac{2}{\pi k\ell^{*}}\ln\left[1+\frac{D(\Delta\nu)\tau}{(s\ell^{*})^2}\cdot\frac{1}{1-2\pi i \Delta\nu \tau}\right]
\label{eqn3}
\end{equation}
with $s\sim 1$, $D_{\mathrm{B}} = (c/n_{\mathrm{eff}})\ell^{*}/2$ the Boltzmann diffusion coefficient, and $\tau$ the lifetime of a photon in our 2D structure. The prefactors $F_{1,2} \leq 1$ in Eq.~(\ref{eqn2}) account for the suppression of measured fluctuations due to the non-zero correlation length of fluctuations of the dielectric function $\ell_{\epsilon}$, and due to the non-zero size of signal collection area $a$.

The first term on the right-hand side of Eq.~(\ref{eqn2}) is the so-called $C_0$ correlation function \cite{Shapiro,Maynard,Page}. It is determined solely by the single scattering \cite{ss}  near the measurement point, it does not depend on $\Delta\nu$ as long as $|\Delta\nu|\ll\nu$ and thus it is often referred to as ``infinite-range''�. Among all the possible scattering events, the single scattering is the fastest one and thus it determines the asymptotic behavior of $\langle C_{\rho}(\Delta\nu)\rangle$ at large detunings $\Delta\nu$. $\ln(2k\ell)/\pi k\ell$ in Eq.~(\ref{eqn2}) represents LDOS variance for the white-noise disorder ($\ell_{\epsilon}\rightarrow 0$). The nonuniversal, disorder-specific nature of $C_0$ is encoded in the function $F_1$ that explicitly depends on the correlation length of disorder $\ell_{\epsilon}$ and suppresses LDOS fluctuations with respect to their value for the white-noise disorder.  The second term on the right-hand side of Eq.~(\ref{eqn2}) is the multiple-scattering contribution to the correlation function decaying with $\Delta\nu$. This term is generated by photons that explore a large area on a time scale exceeding the mean free time $\ell/c$. It encodes the information about multiple scattered photons and controls the decay of $\langle C_{\rho}(\Delta\nu)\rangle$ for small $\Delta\nu$. The function $F_2$ describes the suppression of this term due to the collection of signal from an area of non-zero size in the experiment. The size of the signal collection area $a$ is the same for all our measurements. The suppression factor $F_2$ is evaluated analytically and it decreases with $ka$ \cite{SM}.

\begin{figure}
\includegraphics[width=\columnwidth]{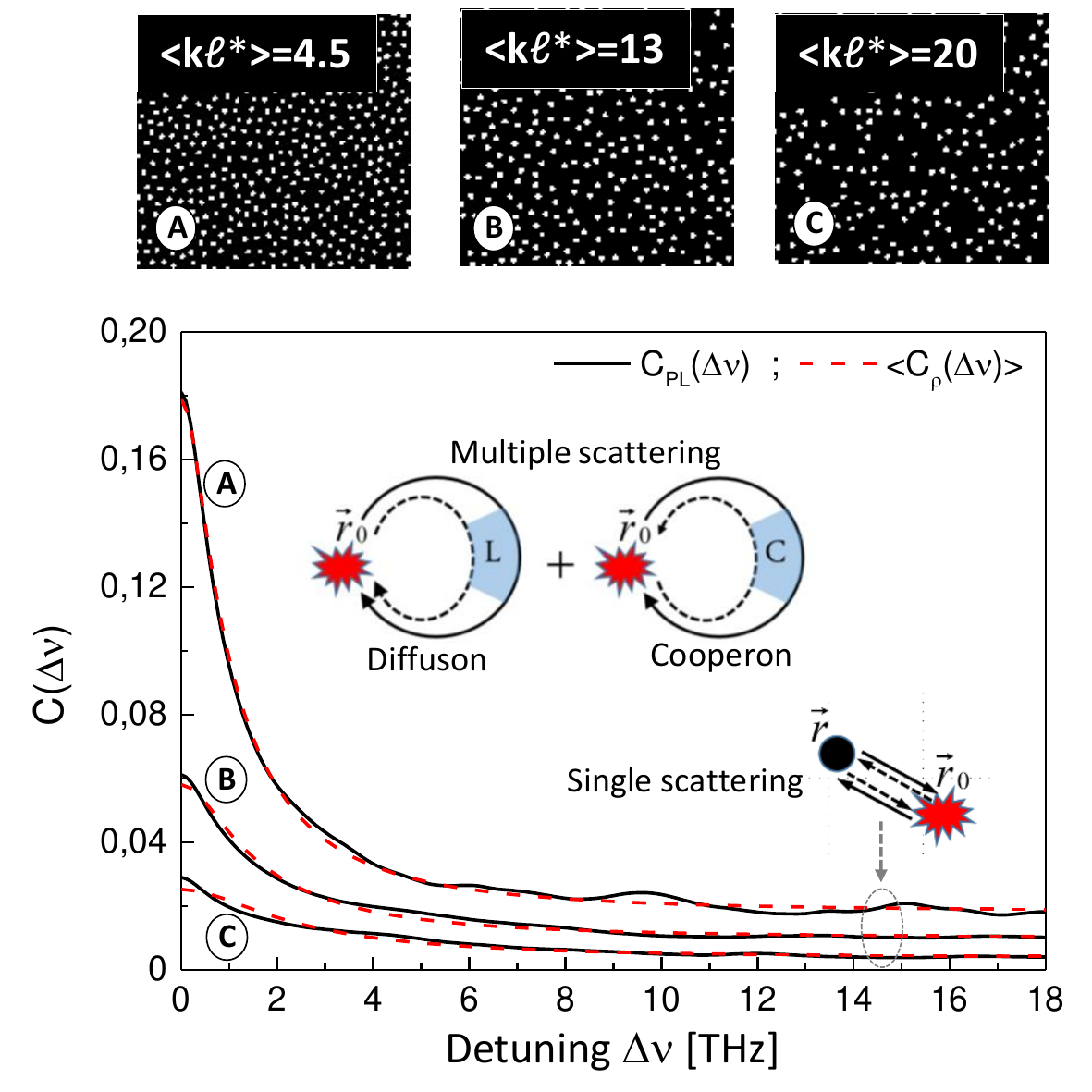}
\caption{\label{fig3} Frequency correlation function of PL (black solid line) and the corresponding theoretical fit with $\langle C_{\rho}(\Delta\nu)\rangle$ (red dashed line) for samples with different scattering strengths $k\ell^{*}$. The inset shows scattering diagrams yielding different contributions to $\langle C_{\rho}(\Delta\nu) \rangle$. The classical (diffuson) and coherent (cooperon) diagrams are multiple scattering contributions that occur on large length and time scales and determine the behavior of $\langle C_{\rho}(\Delta\nu) \rangle$ at small detuning $\Delta\nu$ \cite{SM}. The single scattering is the fastest process that determines the asymptotic tail of $\langle C_{\rho}(\Delta\nu) \rangle$ at large $\Delta\nu$. The top of the image shows the SEM images of samples with different scattering strengths $k\ell^{*}$. }
\end{figure}

To fit the experimental data with Eq.~(\ref{eqn2}) we consider the photon lifetime $\tau$, the nonuniversal suppression factor $F_1$ and $s\sim 1$ as free fit parameters. $\ell$ , $\ell^{*}$ and $D_{\mathrm{B}}$ are estimated using standard approaches from the number density of holes $N$, their average diameter $\langle d_h\rangle$, and the minimum distance $D_{\mathrm{HC}}$ between them \cite{SM}. These quantities can be measured with standard SEM techniques taking advantage of the planarity of our samples.

Figure~\ref{fig3} shows examples of measured $C_{\mathrm{PL}}(\Delta\nu)$ (black solid lines) compared with the theoretical $\langle C_{\rho}(\Delta\nu)\rangle$ (red dashed lines). The three curves correspond to three samples with different degrees of disorder, i.e. different values of $k\ell^{*}$ (samples A, B and C, respectively, shown at the top of Fig.~\ref{fig3}). The decay of $C_{\mathrm{PL}}(\Delta\nu)$ with $\Delta\nu$ is well described by the second term in Eq.~\ref{eqn2} whereas for large $\Delta\nu$, $C_{\mathrm{PL}}(\Delta\nu)$ tends to a limit $C_{\mathrm{PL}}(\infty) > 0$ equal to the first term. The amplitudes of both the short- and the infinite-range contributions to $C_{\mathrm{PL}}(\Delta\nu)$ decrease with $k\ell^{*}$, but the two contributions can be clearly separated in all cases.

\begin{figure}
\includegraphics[width=\columnwidth]{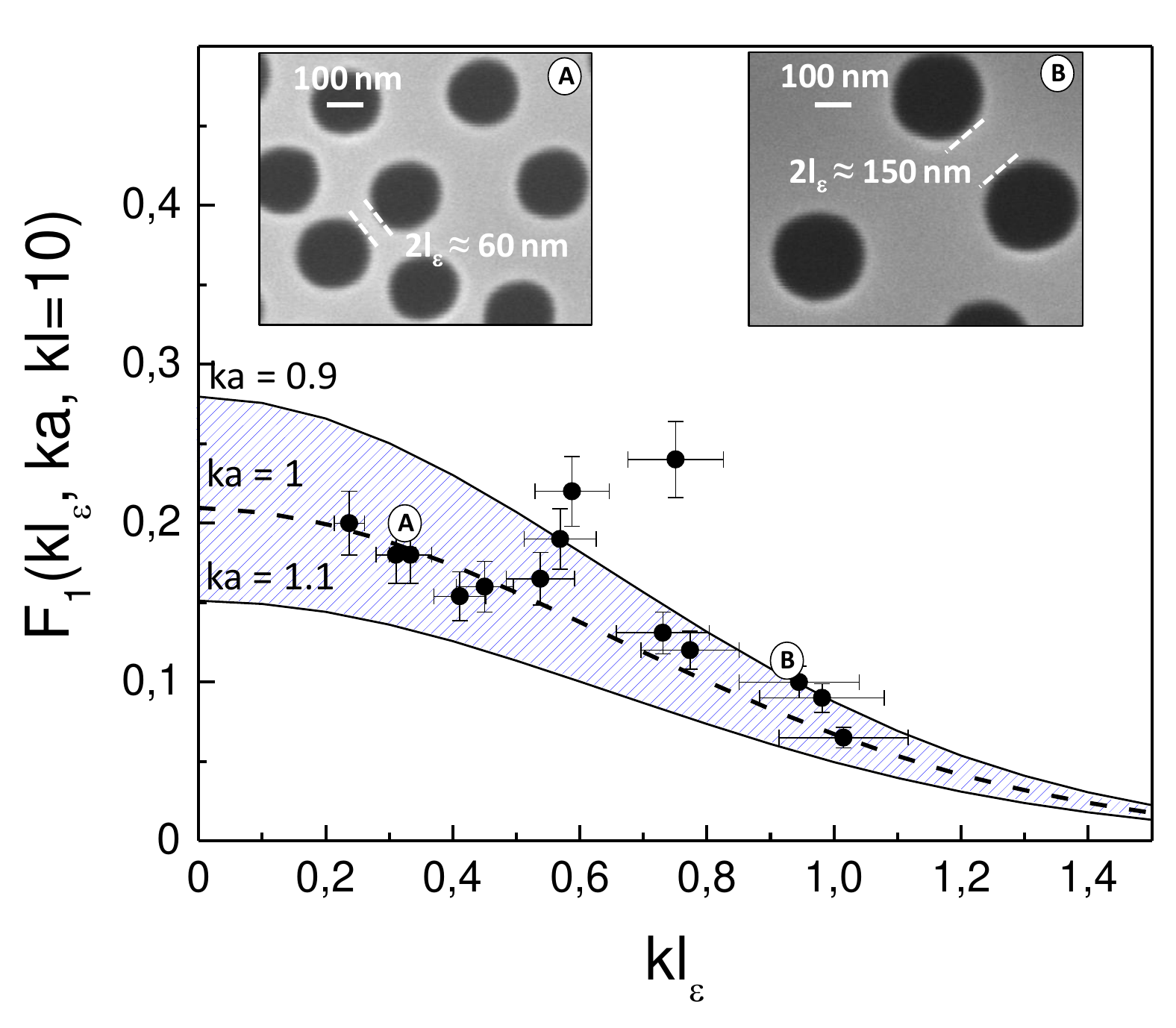}
\vspace{-5mm}
\caption{\label{fig4} Values of $F_1$ (black points) obtained from the fits to the measured $C_{\mathrm{PL}}(\Delta\nu)$ by Eq.~(\ref{eqn2}). The two continuous lines and the dashed line show the behavior of $F_1$ expected from the theory for $ka = 0.9$, 1.1 and 1, respectively, and for $k\ell = 10$ that is typical for the whole set of samples (the dependence on $k\ell$ is very weak). The two insets show the sensitivity of $F_1$ to the average minimum distance between adjacent scatteters $2\ell_{\epsilon}$ \cite{SM}. }
\end{figure}

Figure~\ref{fig4}  shows the best-fit values of the nonuniversal prefactor $F_1$ plotted as a function of $\ell_{\epsilon}$ and compared to a theoretical model in which the correlation function of disorder is assumed to have Gaussian shape [18]. Most of the experimental data fall within the shaded area enclosed between lines corresponding to the two marginal values of $ka$ for our set of samples.

\begin{figure}
\includegraphics[width=\columnwidth]{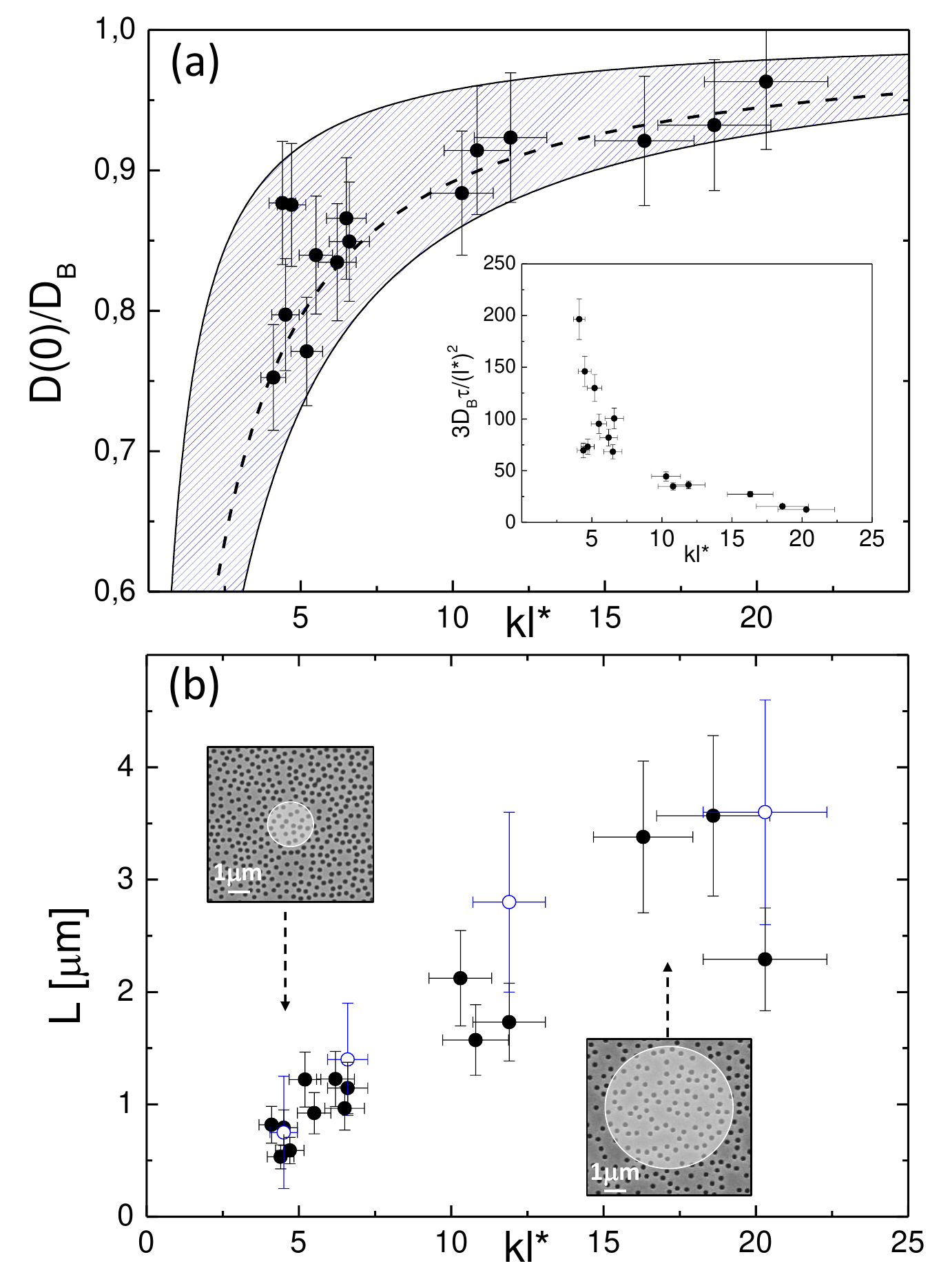}
\vspace{-5mm}
\caption{\label{fig5} (a) Renormalization of the diffusion constant $D(0)$ as a function of $k\ell^{*}$. Full black circles are the values obtained from the theoretical fits to the measured autocorrelation function of PL. The two solid lines show the marginal values of $D(0)/D_{\mathrm{B}}$ for our set of samples. The inset shows the values of the dimensionless parameter $3D_{\mathrm{B}}\tau/\ell^{*2}$ that roughly corresponds to the number of scattering events experienced by a photon before leaving the sample. (b) Characteristic length scale $L$ as a function of $k\ell^{*}$. The four empty circles are the direct measurements of the decay length of photonic modes taken from Ref.~\cite{Riboli}. The two insets show the schematic representation of  $L$ for two values of $k\ell^{*}$. $L$ shrinks with decreasing $k\ell^{*}$.}
\end{figure}

The decay of $C_{\mathrm{PL}}(\Delta\nu)$ is characterized by the lifetime $\tau$ of a photon inside the disordered system, or alternatively the Thouless frequency $\nu_{\mathrm{TH}}=1/\tau$ \cite{Akkermans}. Figure~\ref{fig5}(a) shows that the renormalized diffusion coefficient $D(\Delta\nu = 0)$ calculated using Eq.~(\ref{eqn3}) with our best-fit values of $\tau$ and $s$, goes down to approximately $75\%$ of its Boltzmann value $D_{\mathrm{B}}$ due to Anderson localization effects \cite{Anderson,Abrahams1}. Localization effects are particularly strong in 2D systems and originate from the interference between multiple scattered waves. They become more and more important as the strength of disorder increases, i.e. as $k\ell^{*}$ decreases, and they reduce the value of the diffusion coefficient that eventually goes to zero in the limit of $k\ell^{*}\rightarrow 0$ or $\tau\rightarrow\infty$ \cite{Abrahams2}. The inset of Figure~\ref{fig5}(a) shows the best-fit values of $3D_{\mathrm{B}}\tau/\ell^{*2}$ for our set of samples. This parameter roughly corresponds to the number of scattering events experienced by a photon before leaving the sample. The blue shaded area in Figure~\ref{fig5}(a) is enclosed between the curves corresponding to the two marginal values of $3D_{\mathrm{B}}\tau/\ell^{*2}$. Losses of energy resulting in a finite lifetime $\tau$ of a photon make the 2D material behave as if it was of finite extent $L \sim \sqrt{D_{\mathrm{B}}\tau}/s$. The length scale $L$ encodes the in-plane scattering properties via $D_{\mathrm{B}}$ and the total loss time $\tau$ of the real 3D system. The latter is mainly due to out-of-plane leakage but also accounts for in-plane losses due to the finite sample size. Figure~\ref{fig5}(b) shows that $L$ increases with $k\ell^{*}$ (full black circles); its values are similar to the values of the spatial decay length of photonic modes directly measured in Ref.\ \cite{Riboli} [empty blue circles in Fig.\ \ref{fig5}(b)]. For an infinite 2D disordered system without loss, the latter quantity would be equal to the localization length $\xi$ \cite{Mortessagne}. A separation between contributions of localization and loss to the decay rate of modes in a realistic experiment can be realized by analyzing the statistics of their quality factors \cite{Smolka_NJoP}.

The results of our experiments can be summarized as follows. A QD emits light at a given position $\mathbf{r}$ inside the disordered material and the intensity of emission is measured at the same position $\mathbf{r}$. The single scattering is the fastest mechanism that produces fluctuations of the measured signal with $\mathbf{r}$. This fast contribution gives rise to a large-$\Delta\nu$ tail of $C_{\mathrm{PL}}(\Delta\nu)$. Structural correlations of disorder decrease the amplitude of the signal with respect to its value for uncorrelated (white-noise) disorder but the signal remains well above the noise level and is easily detectable. On the other hand, multiple scattering occurs on longer time scales. It samples a macroscopically large portion of material and have a strong frequency dependence. This mechanism determines the decay of $C_{\mathrm{PL}}(\Delta\nu)$ towards the asymptotic value determined by the single scattering.  Partial averaging of PL fluctuations over the measurement area $S$ reduces both single- and multiple-scattering parts of $C_{\mathrm{PL}}(\Delta\nu)$.

In conclusion, in this work we clearly separate the infinite- and the short-range contributions to the frequency correlation function $C_{\mathrm{PL}}(\Delta\nu)$ of QD photoluminescence. The latter describes the decay of $C_{\mathrm{PL}}(\Delta\nu)$ with $\Delta\nu$ whereas the former accounts for its asymptotic value at large $\Delta\nu$. A direct link between $C_{\mathrm{PL}}(\Delta\nu)$ and the correlation function of LDOS $\langle C_{\rho}(\Delta\nu)\rangle$ is established. Both contributions to $C_{\mathrm{PL}}(\Delta\nu)$ can be understood in the framework of our theoretical model showing that the infinite-range part of $C_{\mathrm{PL}}(\Delta\nu)$ explicitly depends on the disorder correlation length whereas its short-range part is mainly controlled by the renormalization of diffusion due to Anderson localization effects. The separation of physical phenomena behind the two contributions to $C_{\mathrm{PL}}(\Delta\nu)$  and hence to $\langle C_{\rho}(\Delta\nu)\rangle$ allows for efficiently designing a disordered material featuring a particular shape of $\langle C_{\rho}(\Delta\nu)\rangle$. These results pave a way towards designing disordered photonic materials with desired LDOS statistics, opening new perspectives for light-harvesting \cite{Polman,Vynck}, quantum-optics \cite{Sapienza} and light-emission \cite{Wiersma} applications of disordered materials.

F.R. acknowledges P. Sebbah for his help at the initial stage of data analysis, A. Fiore and A. Gerardino for sample fabrication, K. Vynck, S. Vignolini, F. Sgrignuoli and D.S. Wiersma for many fruitful discussions. F.R. acknowledges financial support from the Starting Grant Young Researchers 40600009, University of Trento. S.E.S. acknowledges financial support from the Agence Nationale de la Recherche under grant ANR-14-CE26-0032 LOVE.

%\subsection{\label{sec:level2}Second-level heading: Formatting}
% subsections are not used for PRL papers

%Fig.~\ref{fig:wide} is a figure that is too wide for a single column,
%so instead the \texttt{figure*} environment has been used.
%\begin{figure*}
%\includegraphics{fig_2.eps}% Here is how to import EPS art
%\caption{\label{fig:wide}Use the figure* environment to get a wide
%figure that spans the page in \texttt{twocolumn} formatting.}
%\end{figure*}
%

% avoids incorrect hyphenation, added Nov/08 by SSR
\hyphenation{ALPGEN}
\hyphenation{EVTGEN}
\hyphenation{PYTHIA}

\providecommand{\arccot}{\mathop{\mathrm{arccot}}}

\bibliographystyle{apsrev4-1}
\renewcommand*{\citenumfont}[1]{S#1}
\renewcommand*{\bibnumfmt}[1]{[S#1]}

\setcounter{equation}{0}
\setcounter{figure}{0}

\onecolumngrid
\newpage

\begin{center}
{\large\bf Supplemental material for\\ ``Tailoring correlations of the local density of states in disordered photonic materials''}
\end{center}

\renewcommand{\theequation}{S\arabic{equation}}
\renewcommand{\thefigure}{S\arabic{figure}}
\renewcommand{\bibnumfmt}[1]{[S#1]}
\renewcommand{\citenumfont}[1]{S#1}

\section{\label{sec3} Experimental samples and their characterization}

\subsection{Sample parameters and experimental details}

Our planar samples are characterized by an average hole diameter $\langle d_h \rangle$ ranging from 180 to 250 nm for different samples and an average surface filling fraction $f$ of holes ranging from 0.15 to 0.4 (see Ref.\ \cite{SMRiboli} for details of sample fabrication). The strength of scattering in our samples can be quantified by a product $k\ell^*$. The values of $k\ell^*$ calculated for our samples are almost continuously distributed in a wide range from $k\ell^*=4$ to $k\ell^*=20$, giving us access to both weak ($k\ell^*\gg1$) and strong ($k\ell^*\sim 1$) scattering regimes. The samples are optically activated by inclusion of three layers of InAs quantum dots (QDs) grown by molecular beam epitaxy and embedded in the middle plane of the slab (density of 400--1000 QDs/$\mu$m$^2$, which corresponds to an average inter-dot distance of 30--50 nm). Large inter-dot distances allow us to neglect such interactions between QDs as carrier tunneling (negligible for distances beyond 15 nm \cite{SMWang}) and the dipole-dipole interaction, which becomes important only for distances close to the F{\"o}rster radius (typically 2--9 nm \cite{SMNovotny}). Finally, in our analysis we neglect collective effects mediated by the electromagnetic field which may be important when probing absorption or resonant scattering \cite{SMAgio_PRL,SMCaselli_LSA}, but become negligible for PL lifetimes or intensity measurements.

In a typical experiment, QDs are excited through a dielectric tip of a scanning near-field optical microscope (SNOM, Twinsnom by Omicron) with a 780 nm diode laser (power 60 $\mu$W). We collect the photoluminescence (PL) of QDs through the same tip and analyze its spectrum with the help of a diffraction grating and a 512-pixel linear array of InGaAs photodetectors. For each position $\mathbf{r} = (x,y)$ of the SNOM tip the PL spectrum $I_{\mathrm{PL}}(\mathbf{r},\nu)$ covers a broad wavelength range from $\lambda_0 = 1.15$ $\mu$m to $\lambda_0 = 1.38$ $\mu$m and can be analyzed with a spectral resolution of $0.5$ nm. The spatial resolution of the SNOM tip is $200$ nm, and we scan its position $(x,y)$ through an area of 18 $\mu$m $\times$ 18 $\mu$m.

\subsection{\label{sec4} Structure factor, mean free paths, and correlation length of fluctuations of the dielectric function}

The in-plane scattering and transport mean free paths $\ell$ and $\ell^*$ are calculated thanks to the knowledge of the dielectric function $\epsilon(\mathbf{r})$. The intrinsic planarity of our samples allows us to obtain high-fidelity images of $\epsilon(\mathbf{r})$ and permits to evaluate such structural parameters as the average hole diameter $\langle d_h \rangle$, the filling fraction $f$ of holes, and the structure factor $S(q)$. To fit the experimental data we compute $\ell$  and $\ell^*$ using the following equations \cite{SMConley}:
\begin{equation}
\frac{1}{\ell}=\frac{2N}{\pi kn_{\mathrm{bg}}}\int_0^{\pi}\frac{d\sigma}{d\theta}S\left( 2kn_{\mathrm{eff}}\sin\frac{\theta}{2}\right)d\theta, \quad \frac{1}{\ell^*}=\frac{2N}{\pi kn_{\mathrm{bg}}}\int_0^{\pi}\frac{d\sigma}{d\theta}S\left( 2kn_{\mathrm{eff}}\sin\frac{\theta}{2}\right)(1-\cos\theta)d\theta
\label{eqn13},
\end{equation}
where $N$ is the number density of holes in the sample, $n_{\mathrm{bg}}$ is the effective refractive index of the fundamental TE$_0$ guided mode of the unpatterned slab, and $n_{\mathrm{eff}}$ is the effective refractive index calculated with the porosity of the material taken into account. $S(q)$ is the structure factor describing correlations between holes in the sample. The Boltzmann diffusion coefficient is $D_B=(c/n_{\mathrm{eff}})\ell^*/2$.

\begin{figure}
\includegraphics[width=\columnwidth]{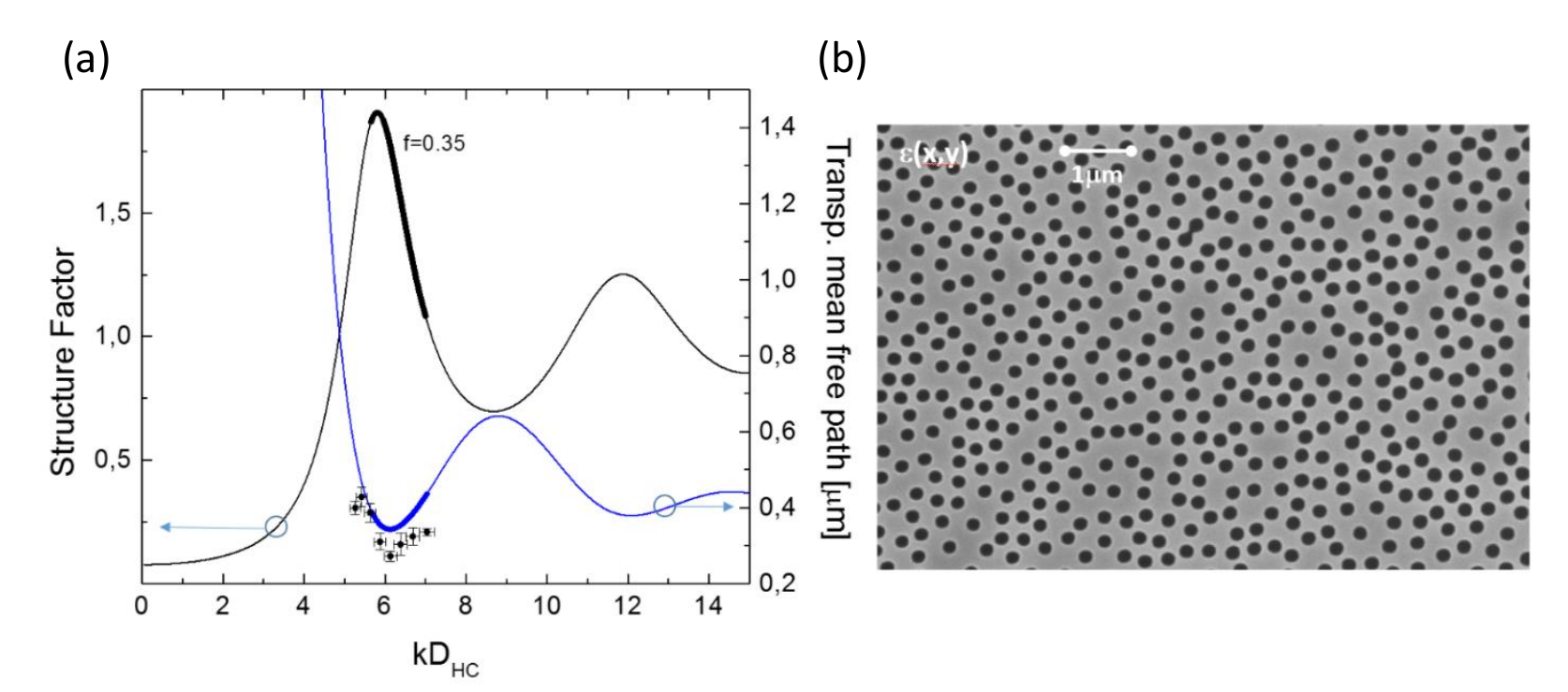}
\caption{\label{figS4} (a) Structure factor (black line) and transport mean free path $\ell^*$ (blue line) of a sample with typical structural parameters ($f=0.35$, $\langle d_h \rangle = 220$ nm, $D_{\mathrm{HC}}=260$ nm). Black circles with error bars show $\ell^*$ evaluated from a numerical FDTD simulation. (b) SEM image of the sample with the structure factor shown in (a).}
\end{figure}

\begin{figure}
\includegraphics[width=0.8\columnwidth]{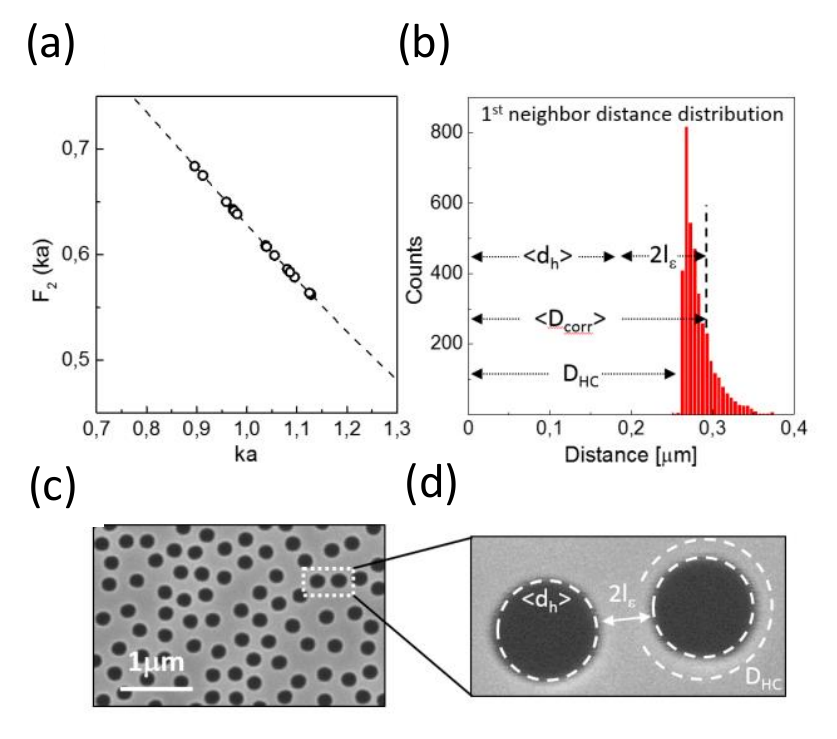}
\caption{\label{figS5} (a) $F_2$ (dashed line) as a function of $ka$ expected from theory. The empty circles represent the values of $F_2$ of our set of samples. (b) Distribution of nearest-neighbor distances and the microscopic structural length scales that define the correlation length $\ell_{\epsilon}$. $D_{\mathrm{HC}}$ is the minimum distance imposed during the random sequential addition design of the sample, $\langle d_h \rangle$ is the average hole diameter, $D_{\mathrm{corr}}$ is the first moment of the distribution. Panels (c) and (d) illustrate the definition of the correlation length of disorder $\ell_{\epsilon}$.}
\end{figure}

Figure \ref{figS4}(a) shows the structure factor $S(q)$ calculated using the coordinates of holes extracted from a SEM image of one of the samples. In the same panel we also show the transport mean free paths calculated using Eq.\ (\ref{eqn13}) (blue line) and numerically using a FDTD code (black dots). Figure \ref{figS4}(b) shows an image of a typical sample. The value of $\ell^*$ that we use for each sample is the average value of the blue curve around the marked region representing our experimental frequency window. The dielectric function $\epsilon(\mathbf{r})$ has been generated with a random-sequential addition (RSA) generator, imposing the hard-core potential with $D_{\mathrm{HC}} = 1.3 \langle d_h \rangle$. Each sample thus is a random packing of cylindrical holes characterized by an average packing fraction that ranges from $p_f=0.2$ to $p_f=0.53$ [$p_f = f(D_{\mathrm{HC}}/\langle d_h \rangle)^2$]. The densest sample ($p_f=0.53$) is close to the jamming limit for RSA, i.e. the random-close-packing limit in which $D_{\mathrm{HC}}=d_h$. This kind of randomness possess a structural length scale $D_{\mathrm{corr}}$ defined as a mean nearest-neighbor distance between two scattering centers [see Fig.~\ref{figS5}(b)]. The correlation length $\ell_{\epsilon}$ is defined as the difference between $D_{\mathrm{corr}}/2$ and the average scatterer radius $\langle d_h \rangle/2$:
\begin{equation}
2\ell_{\epsilon} = D_{\mathrm{corr}}-\langle d_h \rangle.
\label{eqn14}
\end{equation}
$2\ell_{\epsilon}$ thus represents the average minimum distance between the edges of two adjacent holes [see Figs.~\ref{figS5}(c)--(d)] that is a local, microscopic feature of disorder in our samples. The correlation length $\ell_{\epsilon}$ drives the non-universal contribution to LDOS fluctuations determining the amplitude of the suppression factor $F_1$ of the single scattering contribution to $\langle C_{\rho}(\Delta\nu)\rangle$. Figure \ref{figS5}(a) shows the theoretical behavior of the suppression factor $F_2$. The black dashed line is the theoretical behavior also shown in Fig.~\ref{figS2}(b). The empty circles are the values of $F_2$ for the samples investigated in the present work.

\section{\label{sec1} Considerations about the measured photoluminescence and the local density of states}

Many different techniques have been proposed to probe the local density of states (LDOS) of photonic systems. The single-emitter decay-rate experiments \cite{SMLodahl_PRL}, the angular and spectral detection of the electron-induced light emission \cite{SMSapienza_NM}, and the conventional scanning near-field optical microscopy (SNOM) \cite{SMDereux,SMVignolini_APL} are the most prominent examples. Each of these techniques is adapted to probe LDOS under different conditions, i.e. at cryogenic or room temperatures, for different orientations and spatial locations of the light source, with different spatial and spectral resolutions. Measuring LDOS by the conventional scanning near-field optical microscopy has been theoretically discussed and experimentally demonstrated by different authors \cite{SMVignolini_APL,SMGreffet_PRB,SMGreffet_Nature} for various detection schemes, different typologies of the SNOM tip (metal-coated tip or dielectric uncoated tip), or different sample illumination conditions (thermal radiation or quantum light sources). Photoluminescence (PL) signal measured in the near field of a sample in which quantum dots (QDs) are embedded depends, in principle, on many parameters. Intrinsics effects like (i) non-radiative recombination mechanisms, (ii) the local field enhancement factor \cite{SMWenger}, (iii) the absorption scattering cross section of QDs, as well as extrinsic effects, like (iv) perturbations induced by the SNOM tip \cite{SMKoenderink}, (v) small impurities on the sample surface, (vi) the inhomogeneous spatial distribution of quantum dots, considerably affect the magnitude of the measured PL.

The functional relationship between PL and the radiative LDOS is not linear, especially when the non-radiative recombination rate $\Gamma_{\mathrm{NR}}$ is comparable with the radiative one $\Gamma_{\mathrm{R}}$. At room temperature, $\Gamma_{\mathrm{NR}} \gg \Gamma_{\mathrm{R}}$ for In-As quantum dots \cite{SMGurioli_PRB}, and the relation between PL intensity and LDOS can be linearized. The coefficient of proportionality depends on many parameters related to intrinsic and extrinsic effects (see above). On the other hand, if we are interested in the relation between \textit{autocorrelation functions} of the fluctuations PL and LDOS that are normalized by the corresponding averages (see Eq.\ (1) of the main text and Eq.\ (\ref{eqn5}) below), an important requirement is that the cross-correlations between the PL signal and all the other processes (affecting the proportionality coefficient) are smaller than the normalized correlation of LDOS.

Our approach relies on an assumption that the spatial and frequency autocorrelation functions of the near-field PL coincide---to a good accuracy---with the corresponding autocorrelation functions of the local density of optical modes having the electric field in the plane of our quasi-2D samples (for brevity denoted by LDOS in the following). We have tested this assumption in several ways, including a direct quantitative comparison of the PL signal measured in an experiment with the theoretically calculated LDOS, for a photonic crystal cavity fabricated using the same technology as the one used to fabricate disordered samples in the present work. We also use this cavity as a reference to optimize our experimental setup. It should be understood, however, that the relation between PL and LDOS we rely on, is not universal. It holds only for a specific subclass of photonic modes that have sufficiently narrow spectral and spatial widths (e.g., photonic crystal cavity modes or Anderson localized modes). The LDOS of these systems is dominated by resonances corresponding to optical modes with small modal volumes, which we are able to map with high accuracy. Prior to any experimental scan, we make a careful selection of home-made SNOM tips by comparing PL maps with LDOS maps for a photonic crystal cavity that we use as a reference.

\subsection{\label{subsec2} Functional relationship between LDOS and PL signal in our experiments}

\begin{figure}
\includegraphics[width=0.9\columnwidth]{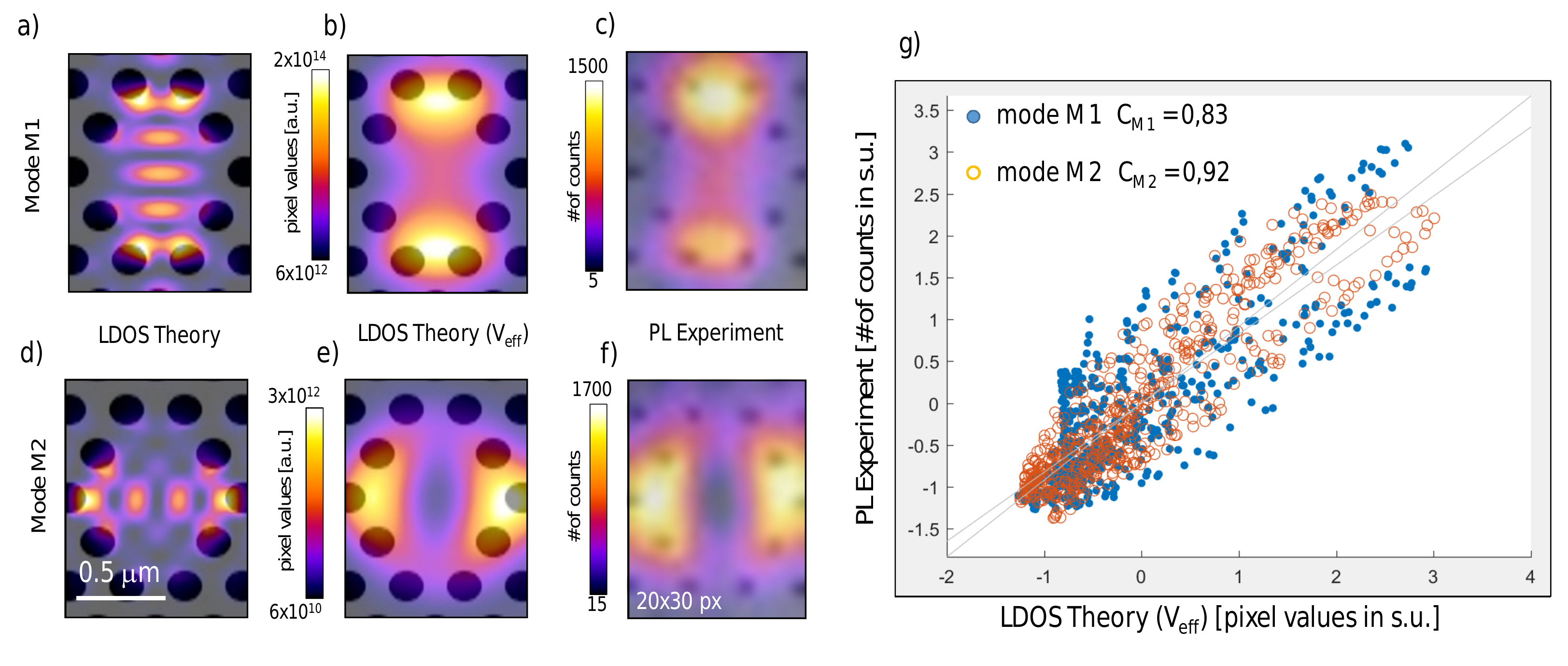}
\caption{\label{figS1_bis} Comparison between calculated mode intensity profiles (a, b, d, e) and measured PL intensity (c, f) for two modes M1 and M2 of a photonic crystal cavity. Panels (a) and (d) show mode intensities in the middle plane of a quasi-2D planar sample (in arbitrary units), which are proportional to LDOS at the corresponding frequencies; panels (b) and (e) show mode intensities integrated over an effective volume $V_{\mathrm{eff}} = S a$, with $S = \pi a^2$ and $a \simeq 100$ nm the effective size of the signal collection area in our experiments; panels (c) and (f) show PL intensity or, more precisely, raw number of photocounts measured by the detector in the experiments. Panel (g) reports PL intensity as a function of the volume-integrated mode intensity (or, equivalently, LDOS) in standardized units $x = (x - \langle x \rangle)/\sigma^2_x$. It demonstrates a linear relationship between PL intensity and LDOS to a good accuracy.}
\end{figure}

By definition, LDOS $\rho(\mathbf{r}, \omega)$ can be expressed via eigenmodes $\psi_n(\mathbf{r})$ of a wave system as
\begin{eqnarray}
\rho(\mathbf{r}, \omega) = \sum\limits_{n} \left| \psi_n(\mathbf{r}) \right|^2 \delta(\omega - \omega_n),
\label{ldosdef}
\end{eqnarray}
where $\omega = 2\pi \nu$.
In a spectral range containing localized eigenmodes, the eigenfrequencies $\omega_n$ are well separated and $\rho(\mathbf{r}, \omega)$ is typically dominated by a single mode: $\rho(\mathbf{r}, \omega_n) \simeq \left| \psi_n(\mathbf{r}) \right|^2$. In its turn, PL intensity is proportional to mode intensity $\left| \psi_n(\mathbf{r}) \right|^2$ as well, as we illustrate in Fig.\ \ref{figS1_bis}. We thus conclude that PL intensity and LDOS are proportional to each other.

Whereas the above reasoning reflects the essence of our experimental approach, the real situation is more complex. LDOS $\rho(\mathbf{r}, \omega)$ that can be assessed via PL in our planar, quasi-2D samples is the local density of states corresponding to transverse electric (TE) modes TE$_0$. These modes have the electric field in the sample plane and exhibit a quasi-uniform intensity distribution as a function of $z$ (the axis perpendicular to the sample plane). They represent the main contribution to the total LDOS in 2D photonic crystal cavities on dielectric slab waveguides \cite{SMIntonti_PRB}. The experimental measurements of PL show a linear dependence on the calculated LDOS, when LDOS is averaged over an effective collection area  $S = \pi a^2$ ($a$ is the spatial SNOM resolution) and are taken at an effective height  $h\simeq a$ above the sample surface  ($a\simeq 100$ nm). In practice, the SNOM tip makes an intrinsic average of the signal over an effective collection volume $V_{\mathrm{eff}} \simeq Sa$  that comprises the emitted fields of many incoherent QDs (roughly 30 QDs within $S$, in samples with $\sim 500$ QDs$/\mu$m$^2$). Figure \ref{figS1_bis} shows a comparison between the measured PL intensity and the calculated LDOS for two cavity modes  \cite{SMIntonti_PRB} that we use as references to test our experimental setup.  Panels (a) and (d) show the calculated LDOS in the slab (TE$_0$ modes), panels (b) and (e) show LDOS integrated over $V_{\mathrm{eff}}$, and panels (c) and (f) show the measured PL maps. The PL maps (c) and (f) nicely fit the shape and the envelope of the numerical calculations in (b) and (e). Only a small intensity unbalance between the two lobes of experimental mode M1 (panel c) reveals artefacts that are unavoidable in real samples such as (i) uncontrolled but weak variations in QDs density or quality, (ii) the presence of small impurities on the surface of the sample or (iii) slight deviations of the structural parameters from the nominal ones. To quantify the degree of similarity between numerically calculated LDOS and experimentally measured PL intensity, we evaluate linear correlation coefficients $C$ between them. The panel (g) of Fig.\ \ref{figS1_bis} shows a scatter plot of PL (values from panels (c) and (f)) versus LDOS (values from panels (b) and (e)), in standardized units (s.u.) The linear correlation coefficients are  $C_{M1} = 0.83$ for the mode M1 and $C_{M2} = 0.92$ for the mode M2, indicating that roughly $80\%$ of total variation in PL can be explained by its linear dependence on LDOS.  The remaining $20\%$ of the total variation of PL is likely to be associated with extrinsic effects (see the discussion below) or other neglected processes that we are not able to control, such as the mutual dependence between LDOS and the collection efficiency of the SNOM tip.

\subsection{\label{subsec3} Perturbation of PL by the SNOM tip}

The SNOM tip can perturb PL of our samples in three ways: spectral shift and broadening of the measured signal, and smearing of the spatial distribution of PL due to a limited spatial resolution (i.e., a wide point spread function) of the tip. All these effects have been taken into consideration or, alternatively, have a negligible impact on the shape and amplitude of the measured frequency autocorrelation $C_{\mathrm{PL}}$. The net effect of the SNOM tip on $C_{\mathrm{PL}}$ is to slightly slow down its short-range decay. The effect is of the order of $\delta \nu / \nu_{\mathrm{TH}}$, where $\delta \nu$ is the typical frequency shift induced by the SNOM tip and $\nu_{\mathrm{TH}}$ is Thouless frequency determining the width of the autocorrelation function $C_{\mathrm{PL}}(\Delta\nu)$. In the following we provide a detailed explanation.

The tip perturbs the local dielectric environment where QD emission and multiple light scattering take place. This perturbation affects the spectrum of PL signal \cite{SMKoenderink}; its importance depends on the shape and size of the apex of the dielectric tip. Previous studies of PL in photonic crystal cavities \cite{SMIntonti_PRB} and disordered systems \cite{SMRiboli} have shown that our typical uncoated dielectric SNOM tips shift the optical resonances of systems under study towards lower frequencies and slightly broaden them. In contrast, they do not perturb significantly the spatial profile of the resonant mode \cite{SMCaselli}. The spectral shift $\delta \lambda$ induced by the dielectric tip is directly proportional to the intensity of the local electric field and inversely proportional to the modal volume of the localized mode. This means that each resonance undergoes a different spectral shift and broadening depending on its modal volume and intensity. The typical tip-induced spectral shift for our disordered photonic systems is $\delta \lambda \simeq 0.4$ nm, corresponding to $\delta \nu \simeq 0.1$ THz \cite{SMRiboli}. This should be compared to the typical width of $C_{\mathrm{PL}}(\Delta \nu)$, which is $\nu_{\mathrm{TH}} \simeq 3$ THz (for $k\ell^{*} = 4$--5, see Fig.\ 3 of the main text). We see that the tip-induced spectral shift $\delta \nu$ is much smaller that the Thouless frequency $\nu_{\mathrm{TH}}$ and can lead only to a weak broadening of $C_{\mathrm{PL}}(\Delta \nu)$. The effects of spectral broadening of resonances have an even smaller impact on the shape of $C_{\mathrm{PL}}(\Delta \nu)$ \cite{SMIntonti_PRB}, also due to the relative small quality factor of localized modes.

The limited spatial resolution of the tip results in a slight suppression of fluctuations of the measured signal with respect to the hypothetical ideal case of point-like detection. In our data analysis, we account for this effect by introducing suppression factors $F_1(k\ell_{\epsilon}, ka, k\ell)$ and $F_2(ka)$ (see Eq.\ (2) and Fig.\ \ref{figS2}). These factors depend, in particular, on the radius $a$ of the signal collection area $S$, determined, in its turn, by the spatial resolution of our SNOM.

\subsection{\label{subsec4}  Polarization of the radiation emitted by QDs}

We experimentally observe that the polarization of the pump light ($\lambda = 780$ nm) does not have any impact on the excitation of QDs. Indeed, the absorption of the pump is due to band-to-band electronic transitions in GaAs and during the carrier energy relaxation, the memory of polarization of the absorbed photon is completely lost. QDs are located in the medial plane of the planar waveguide and the recombination of hole-electron pairs produces light polarized parallel to the slab surface. This is due to the heavy hole character of excitons in QDs \cite{SMStobbe}. The planar waveguide supports four guided modes: TE$_0$, TM$_0$, TE$_1$, and TM$_1$, but only the spatial symmetry and polarization of TE$_0$ mode is compatible with QD emission. Therefore, our experiments probe only LDOS of TE$_0$ modes, which nevertheless represents the main contribution to the total LDOS in photonic crystal cavities and dielectric slab waveguides perforated with air holes.
In the analysis of experimental results, this property is taken into account by considering only wave vectors and scattering and transport mean free paths, corresponding to TE$_0$ modes. For instance, the scattering cross section $\sigma$ and the effective refractive indices $n_{\mathrm{bg}}$ and $n_{\mathrm{eff}}$ entering Eq.\ (\ref{eqn13}) are calculated for the polarization and modal index of the fundamental TE$_0$ mode.

\subsection{\label{subsec5}  Non-radiative recombination processes and the non-universal suppression factor $F_1$.}

The amplitude of the suppression factor $F_1$ depends on the correlation length $\ell_{\epsilon}$. On the other hand, QD non-radiative recombination processes depend on structural inhomogeneities and uncontrolled parasitic recombination channels. We have verified that this last mechanism does not affect the analysis that we apply to determine $F_1$. Indeed, despite the fact that at room temperature, the non-radiative decay rate $\Gamma_{\mathrm{NR}}$ of our QDs is much larger than the radiative one $\Gamma_\mathrm{R}$  \cite{SMGurioli_PRB}, an important requirement is that the fluctuations of $\Gamma_{\mathrm{NR}}$ and the cross-correlation between $\Gamma_{\mathrm{NR}}$ and LDOS are negligible with respect to the intrinsic LDOS fluctuations and correlations, respectively. This is indeed a typical situation that we encounter for photonic crystal cavities. The spectrum of PL measured in such cavities exhibits well-defined peaks due to localized cavity modes (typically, $10^4$ photocounts over a background of 100 counts). When we slightly move away from the cavity but still remain inside the photonic material and the frequency band gap, the PL signal drops down. Still, by moving the SNOM tip from point to point (thus investigating different positions) we observe fluctuations of the signal smaller than 10\%. This is an upper estimation of fluctuations induced by possible non-radiative recombination processes. We therefore believe that the observed behavior of the non-universal suppression factor $F_1$ is not affected by non-radiative recombination mechanisms.

\section{\label{sec1} Theoretical model for LDOS correlation function}

As we discussed above, the correlation functions of PL intensity $C_{\mathrm{PL}}(\Delta\nu)$ and of LDOS $C_{\rho}(\Delta\nu)$ could be assumed roughly equal if the SNOM were measuring a signal from a single QD. In reality, the SNOM tip collects emissions from many QDs inside an area $S$ around $\mathbf{r}$ and thus $C_{\mathrm{PL}}(\Delta\nu)$ is related to the frequency and spatial correlation function of LDOS $C_{\rho}(\Delta\mathbf{r},\Delta\nu)$ via
\begin{equation}
 C_{\mathrm{PL}}(\Delta\nu)= \langle C_{\rho}(\Delta\nu)\rangle = \frac{1}{S^2}\int_S d^2\mathbf{r'}\int_S d^2\mathbf{r''}C_{\rho}(\mathbf{r'}-\mathbf{r''},\Delta\nu).
\label{eqn1}
\end{equation}
Therefore, our experiment gives access to the correlation function of LDOS averaged over a small area $S$.

\subsection{\label{sec2}Correlation function of LDOS}

LDOS at a point $\mathbf{r}$ at frequency $\omega = 2\pi\nu$ is related to the imaginary part of the Green's function $G(\mathbf{r},\mathbf{r},\omega)$ of the Helmholtz equation \cite{SMAkkermans}:
\begin{equation}
\rho(\mathbf{r},\omega) = -\frac{2\omega}{\pi c^2} \mathrm{Im} G(\mathbf{r},\mathbf{r},\omega)
\label{eqn2}.
\end{equation}
The Green's function $G(\mathbf{r},\mathbf{r'},\omega)$ obeys
\begin{equation}
 \left\{ \nabla^2+k^2 [ 1+\delta\mu (\mathbf{r}) ] \right\} G(\mathbf{r},\mathbf{r'},\omega) = \delta(\mathbf{r}-\mathbf{r'})
\label{eqn3},
\end{equation}
where  $k = \sqrt{\langle \epsilon \rangle}\omega /c$, $\langle \epsilon \rangle$ is the average value of the dielectric constant $\epsilon$ in the disordered medium, $c$ is the vacuum speed of light, and  $\delta\mu (\mathbf{r})=[\epsilon(\mathbf{r})-\langle \epsilon \rangle]/\langle \epsilon \rangle$ is the relative fluctuation of  $\epsilon$. The average of the Green's function over fluctuations of  $\delta\mu (\mathbf{r})$ in 2D is \cite{SMAkkermans}
\begin{equation}
 \langle G(\mathbf{r},\mathbf{r'},\omega) \rangle = -\frac{i}{4}H_0^{(1)}\left[ \left( k + \frac{i}{2\ell}\right)|\mathbf{r}-\mathbf{r'}|\right],
\label{eqn4}
\end{equation}
where $\ell$ is the scattering mean free path.

\begin{figure}
\includegraphics[width=0.8\columnwidth]{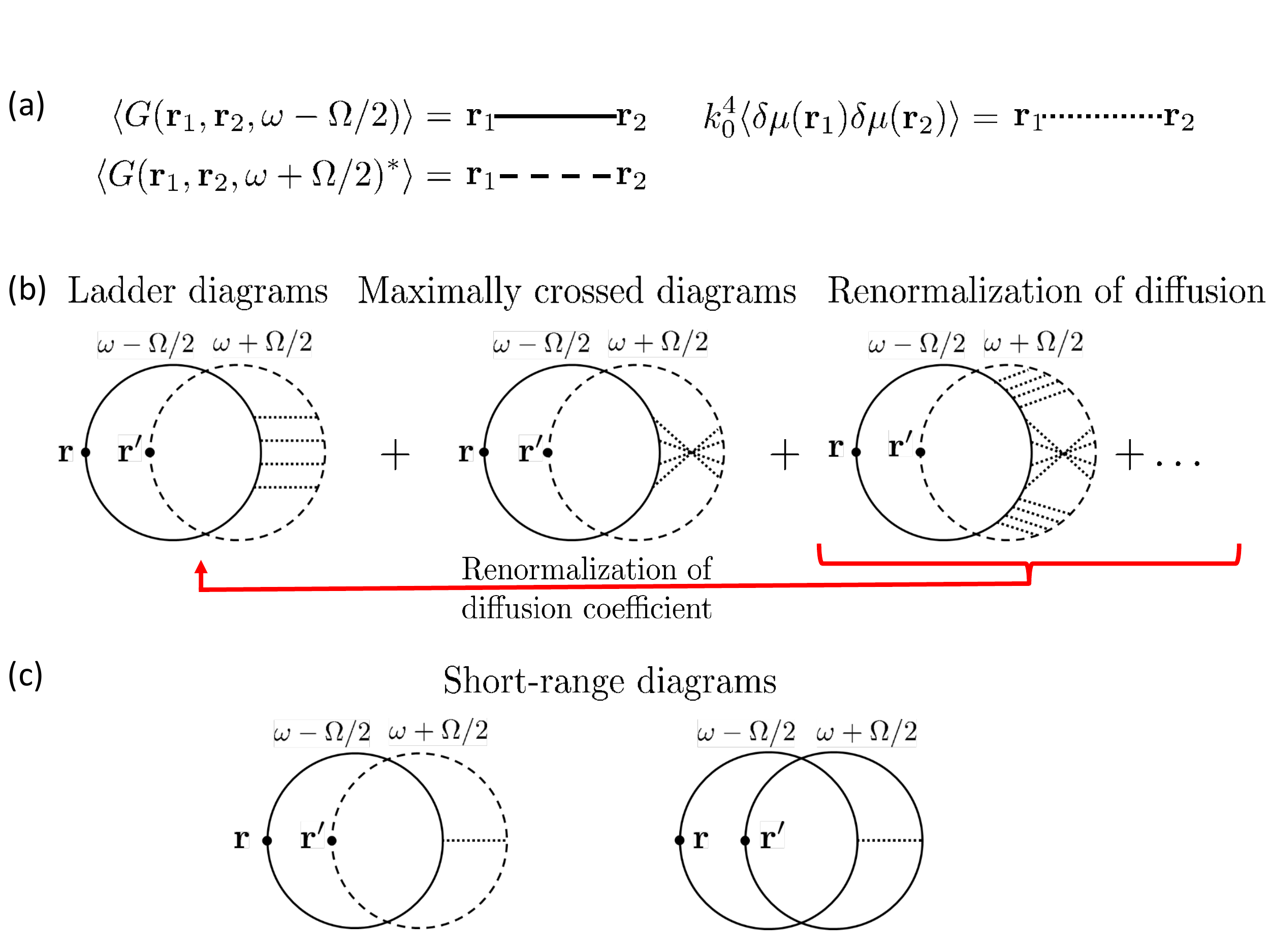}
\caption{\label{figS1} Diagrammatic representation of the correlation function of LDOS. (a) Definition of diagrammatic notations. (b) Universal and (c) nonuniversal contributions to the correlation function of LDOS fluctuations. }
\end{figure}

To compute the correlation function of LDOS
\begin{equation}
C_{\rho}(\Delta\mathbf{r},\Omega) = \frac{\langle \delta\rho (\mathbf{r}+\Delta\mathbf{r}/2,\omega+\Omega/2)\delta\rho (\mathbf{r}-\Delta\mathbf{r}/2,\omega-\Omega/2)\rangle}{\langle \rho(\mathbf{r},\omega)\rangle^2}
\label{eqn5}
\end{equation}
we use Eq.~(\ref{eqn2}) and the standard perturbative diagrammatic techniques to average products of Green's functions \cite{SMAkkermans}. The resulting diagrams are illustrated in Fig.~\ref{figS1} and yield $C_{\rho}$ as a sum of two distinct contributions. First, the universal contribution results from the diagrams of Fig.~\ref{figS1}(b):
\begin{equation}
C_{\rho}^{(U)}(\Delta\mathbf{r},\Omega) = f^2(\Delta r) \mathrm{Re} \left[\frac{D_B}{D(\Omega)}-1\right],
\label{eqn6}
\end{equation}
where
\begin{equation}
 f(\Delta r) = \frac{1}{1-\frac{2}{\pi} \arccot (2k\ell)} \mathrm{Re} H_0^{(1)}\left[ \left( k + \frac{i}{2\ell}\right)\Delta r\right]
\label{eqn7}
\end{equation}
and the renormalized diffusion coefficient $D(\Omega)$ will be defined in the next section.

\begin{figure}
\includegraphics[width=0.49\columnwidth]{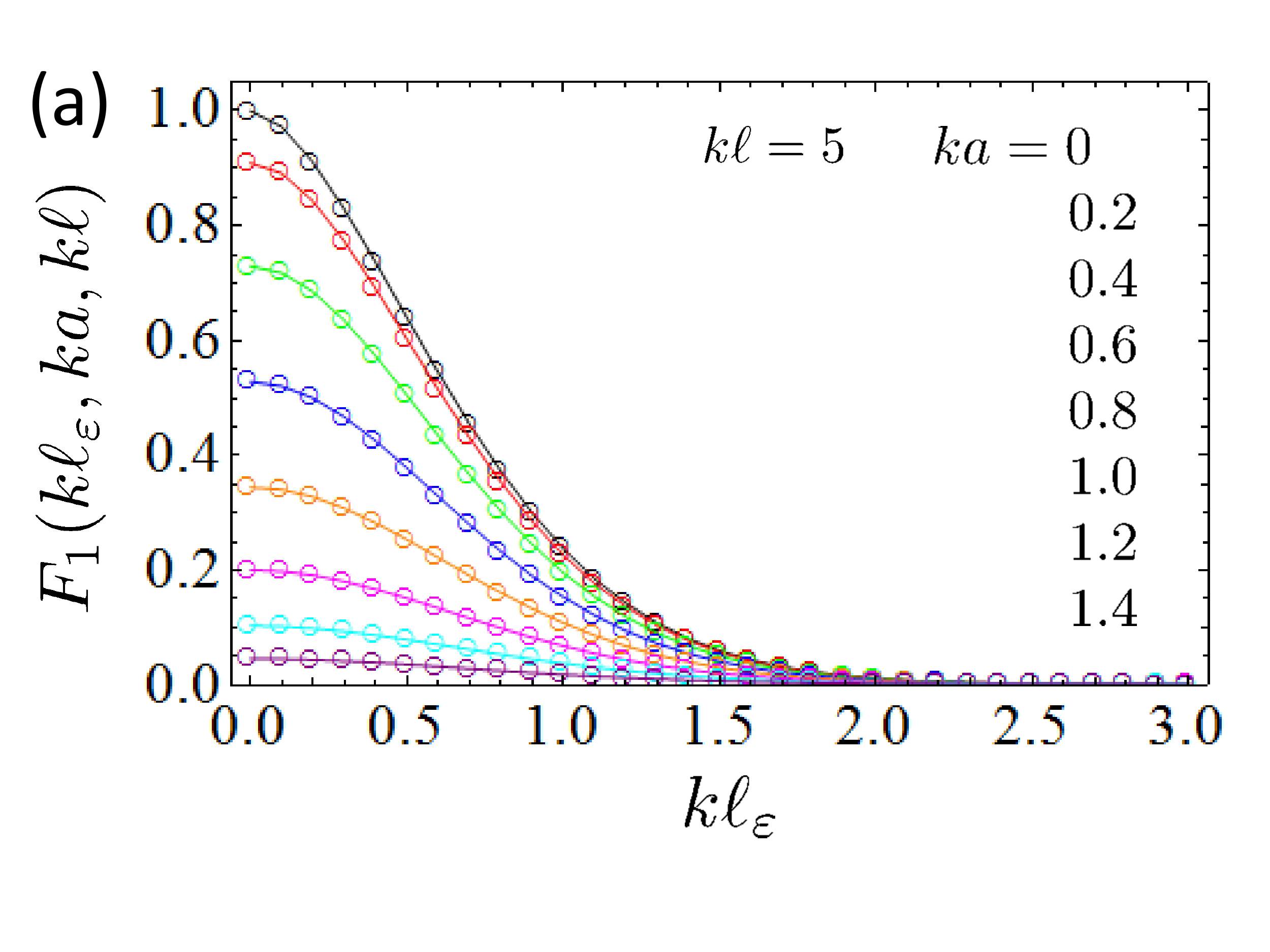}
\includegraphics[width=0.49\columnwidth]{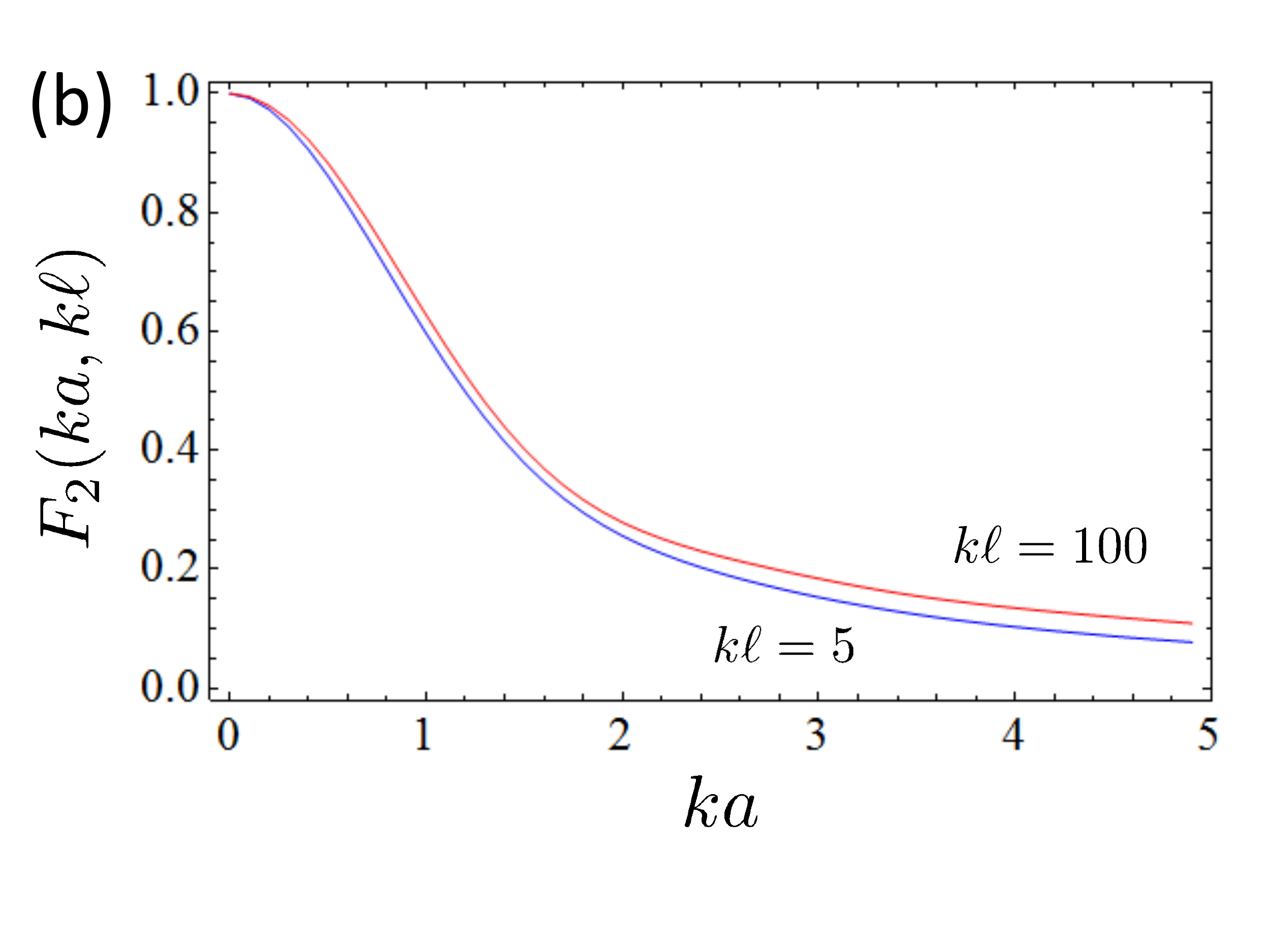}
\caption{\label{figS2} Factors describing the suppression of nonuniversal (a) and universal (b) contributions to the frequency correlation of PL intensity due to the finite signal collection area $S =\pi a^2$ and non-zero disorder correlation length $\ell_{\epsilon}$, for a single value of $k\ell = 5$ in (a) and for two different values $k\ell = 5$ and $100$ in (b). }
\end{figure}

The second, nonuniversal contribution results from the calculation of short-range diagrams of which examples are shown in Fig.~S1(c), and can be written as an integral to be calculated numerically:
\begin{eqnarray}
C_{\rho}^{(NU)}(\Delta\mathbf{r},\Omega) &=& \frac{k}{\ell}\int d^2\mathbf{r}\int d^2\mathbf{r'}h(\mathbf{r}-\mathbf{r'}) \mathrm{Re} H_0^{(1)}\left[ \left( k + \frac{i}{2\ell}\right)r\right]\mathrm{Im} H_0^{(1)}\left[ \left( k + \frac{i}{2\ell}\right)r\right]  \nonumber \\
&\times& \mathrm{Re}H_0^{(1)}\left[ \left( k + \frac{i}{2\ell}\right)|\Delta\mathbf{r}-\mathbf{r'}|\right]\mathrm{Im} H_0^{(1)}\left[ \left( k + \frac{i}{2\ell}\right)|\Delta\mathbf{r}-\mathbf{r'}|\right],
\label{eqn8}
\end{eqnarray}
where the function $h$ depends on the form of the correlation function of the fluctuations of $\delta\mu(\mathbf{r})$. For Gaussian correlation,
\begin{equation}
 h(\mathbf{r}-\mathbf{r'}) = \frac{1}{\sigma_{\mu}^2\pi\ell_{\epsilon}^2}\langle \delta\mu(\mathbf{r})\delta\mu(\mathbf{r'})\rangle = \frac{1}{\pi\ell_{\epsilon}^2}\mathrm{exp}\left[-\frac{|\mathbf{r}-\mathbf{r'}|^2}{\ell_{\epsilon}^2} \right],
\label{eqn9}
\end{equation}
where $\ell_{\epsilon}$ is the correlation length of $\delta\mu(\mathbf{r})$.
We obtain Eq.~(2) of the main text for the frequency correlation function of PL intensity from $C_{\rho}(\Delta\mathbf{r},\Omega) =C_{\rho}^{(U)}(\Delta\mathbf{r},\Omega) + C_{\rho}^{(NU)}(\Delta\mathbf{r},\Omega)$ using Eq.~(\ref{eqn1}) and $\Omega = 2\pi\Delta\nu$. The resulting suppression factors $F_1$ and $F_2$ are shown in Fig.\ \ref{figS2}. Their dependence on the scattering length is very weak, at least for $k\ell > 5$, and can be neglected within the accuracy of our analysis.

\subsection{\label{sec2}Renormalization of the diffusion constant}

An infinite series of diagrams with crossed diagrams inserted in between two series of ladder diagrams as in the last diagram of Fig.~~\ref{figS1}(b) can be summed up in the same way as it is done when the transport through a disordered medium is calculated \cite{SMVollhardt}. This leads to the renormalization of the diffusion coefficient to be used in the calculation of the sum of ladder and crossed diagrams in the diffusion approximation: $D_B\rightarrow D(\Omega)$ \cite{SMVollhardt,SMCherroret}. The equation for $D(\Omega)$ is
\begin{equation}
\frac{D_B}{D(\Omega)}=1+\frac{2}{\pi\rho_0}P(\mathbf{r},\mathbf{r},\Omega),
\label{eqn10}
\end{equation}
where $\rho_0=\omega/2\pi c^2$ and $P(\mathbf{r},\mathbf{r'},\Omega)$ is the intensity Green's function obeying a diffusion equation
\begin{equation}
\left[ -i\Omega+\frac{1}{\tau}-D(\Omega)\nabla^2 \right]P(\mathbf{r},\mathbf{r'},\Omega) = \delta(\mathbf{r}-\mathbf{r'}).
\label{eqn11}
\end{equation}

\begin{figure}
\includegraphics[width=0.6\columnwidth]{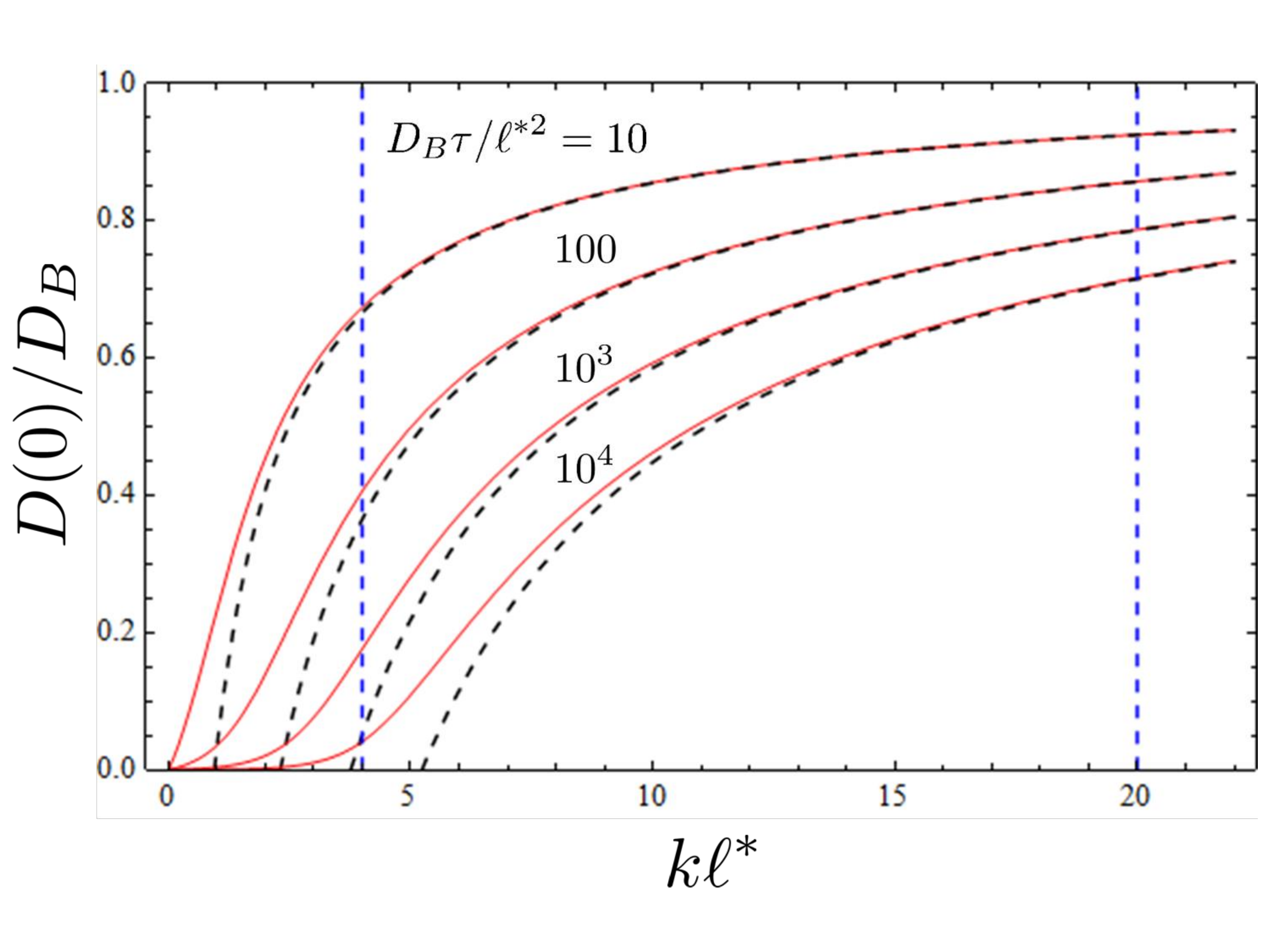}
\caption{\label{figS3} Renormalization of the stationary ($\Omega = 0$) diffusion coefficient as a function of disorder strength  $k\ell^*$ for different values of the photon lifetime $\tau$. Solid lines are exact numerical solutions of Eq.~(\ref{eqn12}) with $s=1$. Dashed lines show first-order perturbative solutions valid for $k\ell^*\gg1$. Vertical dashed lines delimit the approximate region of $k\ell^*$ to which our samples belong.}
\end{figure}

Following the approach of Ref. \cite{SMCherroret}, we obtain from Eqs.~(\ref{eqn10}) and (\ref{eqn11}) the following closed equation for $D(\Omega)$ in a 2D disordered medium:
\begin{equation}
\frac{D(\Omega)}{D_B}=1-\frac{2}{\pi k\ell^*}\ln \left[ 1+ \frac{D(\Omega)\tau}{(s \ell^*)^2}\cdot\frac{1}{1-i\Omega\tau}\right],
\label{eqn12}
\end{equation}
where $s\sim 1$ is a numerical constant determining the precise position of the large-momentum cut-off $q_{\mathrm{max}}=1/s\ell^*$ needed to regularize the divergence of $P(\mathbf{r},\mathbf{r},\Omega)$ in Eq.~(\ref{eqn10}). This nonlinear algebraic equation can be easily solved numerically for any disorder strength $k\ell^*$ or, alternatively, perturbative solutions in any order of $1/k\ell^*$ can be obtained for $k\ell^* \gg 1$. We used such solutions to fit our data in the main text. We present a comparison of exact and perturbative solutions of Eq.~(\ref{eqn12}) for $\Omega = 0$ in Fig.~\ref{figS3}.

\begin{thebibliography}{99}
\bibitem{ss}
We use the term ``single scattering'' for brevity. In reality, light may be multiply scattered but only a single common scattering event is allowed between two waves returning to the source point (see the inset of Fig.~\ref{fig3}). This effectively restricts the relevant area around the source to $\sim\ell^2$.
\end{thebibliography}
\end{document}